# When Does it Pay Off to Learn a New Skill? Revealing the Complementary Benefit of Cross-Skilling

Version: February 11th, 2021[1]

Fabian Stephany✦

✦ Oxford Internet Institute, University of Oxford, UK,
Humboldt Institute for Internet and Society, Berlin,
fabian.stephany@oii.ox.ac.uk

## Abstract

This work examines the economic benefits of learning a new skill from a different domain: cross-skilling. To assess this, a network of skills from the job profiles of 14,790 online freelancers is constructed. Based on this skill network, relationships between 3,480 different skills are revealed and marginal effects of learning a new skill can be calculated via workers' wages. The results indicate that learning in-demand skills, such as popular programming languages, is beneficial in general, and that diverse skill sets tend to be profitable, too. However, the economic benefit of a new skill is individual, as it complements the existing skill bundle of each worker. As technological and social transformation is reshuffling jobs' task profiles at a fast pace, the findings of this study help to clarify skill sets required for designing individual reskilling pathways. This can help to increase employability and reduce labour market shortages.

[1] Acknowledgement: This work has greatly benefited from the exchange with Vili Lehdonvirta on task automation, reskilling strategies, and online labour markets. Likewise, the author thanks Fabian Braesemann for his discussions on network perspectives to technologies, as well as, the participants of the 2018 UNDP Innovation Days in Istanbul.

## Introduction

This paper asks: When does it pay-off to learn something new? It examines the economic benefits of cross-skilling; the process of learning a new skill from a different skill domain. In doing so, this work leverages data of 14,790 online freelancers and their skill portfolios to create a network in which 3,480 skills are connected if they are jointly held by the same worker. With the example of popular programming languages, the findings of this study show that learning one single skill can add more than 50 percent on the average worker wage. The results suggest that acquiring skills from different domains could be of even higher economic value. However, the economic value of cross-skilling largely depends on the composition of the existing skill bundle. The value of learning something new is clearly complementary to what we already know.

The work is motivated by the rapidly changing composition of occupations due to task automation (Frey & Osborne, 2017; Acemoglu & Autor, 2011), resulting in the paradoxical situation of simultaneous unemployment and labour shortage (Autor, 2015). Professional service or admin, white collar, jobs are particularly exposed to this trend that is "hollowing-out" the middle employment spectrum (Baldwin & Forslid 2020). A conventional policy response has been to align national education systems with changing labour market demand. This reply is increasingly ineffectual as technological and social transformation outpaces national education systems (Collins & Halverson, 2018). Workers have to some extent begun to assume greater personal responsibility for reskilling, via skill-based online training (Allen & Seaman, 2015; Lehdonvirta, Margaryan, & Davies, 2019). However, often the economic benefits and costs of reskilling strategies are unclear, as they are highly individual, and precise skill requirements for mastering emerging technologies, such as AI or Big Data, remain opaque (De Mauro et al., 2018).

This work aims to overcome re-skilling uncertainties by assessing the economic benefit of cross-skilling strategies and sketching valuable training pathways, complementary to existing individual skill sets. Furthermore, the empirical relationship of digital skill sets will help to establish a common taxonomy to be used by policy makers, education providers, and recruiters, so that job market mismatches can be reduced. The know-how of this real-time and market data-driven evaluation can be developed into a tool for job market entrants and targeted re-education campaigns. Globally, the value of such a "cross-skilling compass", as presented with a first interactive online prototype for this project[2], could be highest in regions where traditional education infrastructure is lagging behind.

[2] http://www.aiskills.net/

The remainder of this study is organised as follows: In the next section, a literature review embeds the work into discussions on automation of tasks and personalisation of training. Section two illustrates the approach of the work and highlights the importance of skill diversity. Section four presents the data collected and the methods. Section five summarises the results and section six concludes with policy implications and possible extensions of the work.

## Background

### *Automation of Tasks*

The periodic warning that automation and new technologies are going to terminate large numbers of jobs is a recurring theme in economic literature (Frey & Osborne, 2017; Brynjolfsson & McAfee, 2014; Acemoglu & Autor, 2011). A popular early historical example is the Luddite movement of the early 19th century: A group of textile artisans in England protested the automation of their industry by seeking to destroy some of the machines. In contrast to recurring fears of mass unemployment, current literature shows that the (digital) technology revolution, rather automates tasks than vanishing entire occupations (Autor, 2015).

In this process, technological and social transformation change the skill composition of professions (Acemoglu & Autor, 2011). The work that is thereby eliminated has different skill requirements than the newly created jobs, resulting in the paradoxical situation of simultaneous unemployment and labour shortage (Autor, 2015). As the pace of technological and social change accelerates, the skills gap grows rapidly (Milano, 2019). The professional service and admin sector is expected to be at high risk of automation (Frey and Osborne, 2017). This "hollowing out of the middle" spectrum of the labour market (Baldwin & Forslid 2020) turns white collar office workers into the "weavers of the 21st century", most in need of updating their skill sets. History suggests that the skills gap, even more so than the elimination of jobs per se, causes heightened economic inequality (Card & DiNardo, 2002) and retards firm growth (Krueger & Kumar, 2004) during times of technological and social transformation.

More fundamentally, the very notion of occupations is increasingly problematic in large sectors of the economy. The contemporary notion of occupations arose from the industrial revolution, as mass production required large numbers of workers with uniform bundles of skills (Featherman & Hauser, 1979). But today's knowledge workers strive to build unique specialisms and combinations of skills that differentiate them from other workers (Hendarman & Tjakraatmadja, 2012). Even low-end service workers can end up developing extremely heterogeneous skill sets, because they cobble together incomes from idiosyncratic combinations of gigs ranging from coffee serving to Uber driving (Fuller, Kerr, & Kreitzberg, 2019). Tracking labour demand in terms of occupations assumed to consist of uniform bundles of skills therefore fails to produce the kind of information that individual, corporate, and national decision makers need to successfully overcome the skills gap.

A conventional policy response to closing the skill gap has been to align national education systems with changing labour market demand. This response is increasingly ineffectual as technological and social transformation outpaces national education systems (Collins & Halverson, 2018). Workers and education providers alike are uncertain which new, often digital, skill is the first step to an individual and sustainable re-skilling strategy. Large employers are likewise struggling to keep their workforces' skills up to date (Illanes et al., 2018). In response to the uncertainties of traditional re-skilling, researchers have begun to understand occupations as interchangeable bundles of skills or tasks (Bechichii et al., 2018). Dawson et al. (2020), for example, leverage large scale online data on Australian job profiles to create transition pathways from occupations with high risk of automation to professions with lower risks of technological worker displacement. Likewise, workers have to some extent begun to assume greater personal responsibility for reskilling themself, via online courses, distance education tools, and entrepreneurial approaches to work (Allen & Seaman, 2015). This trend is amplified as the COVID-19 pandemic tightens economic budgets and forces workers into individual and remote reskilling.

## *Personalisation of Training*

Similar to the reshuffling of task compositions, digital technologies have enabled a process that has become a defining paradigm of the digital economy: rebundling (McManus et al., 2018). First, in the early days of the Internet, download platforms, at times operating illegally, allowed music lovers to access songs individually without having to acquire the artist’s entire album. The single item (song) was unbundled from the original bundle (album). Later, at a second stage, streaming platforms, like Spotify, reversed the trick by allowing the (re)bundling of previously unrelated items. Users could listen to songs from different artists for one single price. The mastery of this strategy has made digital entertainment companies superstars firms (Eriksson et al., 2019).

In music (Dabager et al., 2014), broadcasting (Hoehn & Lancefield, 2003) or gaming (McManus et al., 2018), things that have been unbundled rarely remain that way. The economic benefit of individualised rebundling is too strong. Similarly, this paradigm has affected the way we learn new skills. At first, in the debundling phase, digital technologies allowed education providers to provide topical online courses (Wulf et al., 2014). At a later stage, platforms like Coursera or DataCamp performed the rebundling and offered a whole set of topical courses for a single price (Bates, 2019). The acquisition of individual skills (programming in Python) has been detached from its original domain of training (studying informatics). However, despite advances in personalised reskilling, a sizable skill gap persists on the labour market.

Current approaches to addressing the skills gap are based on predicting demand for entire occupations or at best for abstract skills such as social skills or creativity. But in many occupations, technological and social transformation is leading the concrete skills that make up the occupation to change regularly and decisively. Nursing has been transformed by successive generations of electronic health record systems, clinical decision support systems, and diagnostic technologies (Adams et al., 2000). Web application development has rapidly rotated from perl to PHP to Ruby to Python to other development platforms (Purer, 2009). Firms and workers who fail to reskill while there is still demand for their skills risk dropping out of the market entirely once demand tips.

Most recent research shows that just-in-time skills development, motivated by the demands of the work at hand, or by perceived market shifts, has emerged, as formal training courses are unaffordable for workers who can't take time off paid work (Kester et al., 2006). In addition, cultural aspects in traditional STEM education, for example, still hinders female participation, despite efforts to alter it (Kahn & Ginther, 2017). Instead research shows that independent professionals, including women, prefer informal, digital, social learning resources like Stack Overflow and tutorial videos to develop new skills (Yin et al., 2018).

Newest findings show that independent IT professionals today develop new skills incrementally, adding closely related skills to their existing portfolio (Lehdonvirta, Margaryan, & Davies, 2019). Their work examines the skill development of freelancers on online labour platforms. Indeed, online freelance platforms might have become early "laboratories" for the de- and rebundling of incremental skills. It could be argued that, for work, freelance platforms, such as UpWork[3], have become what Spotify is for music: They allow freelancers to jointly sell previously detached skill components for one hourly price. The role of the Data Scientist is a prime example of how the rebundling of skills from different domains, i.e., visualisation, programming, and statistics, is an economically profitable offer. The work by Anderson (2017) confirms that diverse rebundles of skills from different domains are profitable in general.

In this situation of rapidly changing market dynamics, systematic oversight is key. However, individuals often lack foresight into which skills are rising, which skills are most valuable and which skills their existing portfolio is complementary to. They get locked into path dependencies that may result in dead ends that prevent them from re-skilling into new areas (Escobari, Seyal, & Meaney, 2019).

[3] UpWork is arguably the most popular online freelance platform internationally (Kässi & Lehdonvirta, 2018).

## The Complementary Value of Learning a New Skill

From an economic perspective, we balance costs and benefits, when deciding to learn a new skill (Gathmann & Schönberg, 2010). Hoping that a new skill will bring us a more income or a better job, we want to limit our effort in acquiring this new capability. In times of digital automation, this re-skilling consideration can be extremely difficult, as often both costs and benefits of newly emerging skills are unclear. Furthermore, the cost-benefit evaluation of learning something new is an individualistic process in general, as costs and benefits are complementary (Anderson, 2017; Allinson & Hayes, 1996). One and the same skill can inflict different costs and leverage different benefits depending on the worker's existing skill set that it is supposed to complement. For reskilling, these interdependencies make the evaluation of skills a complex and highly individual endeavour with no single best choice for everybody. In fact, estimating the individual complementarities of learning a new skill is THE key for finding sustainable reskilling pathways in times of technological change.

In light of the individual complementarities in learning, the rapid reshuffling of occupational profiles, and the failed attempts to develop farsighted re-skilling strategies, this work proposes an economic evaluation of skilling pathways with, at least, the following four goals:

1) Develop an endogenous relationship of skills.
2) Find a method of economic evaluation for learning a new skill.
3) Reveal the complementary benefits of learning a new skill.

In pursuing these goals, the study can rely on previous data-driven approaches to assess skill and human capital evaluation. Traditionally, measures of human capital rely on the count years of experience, training, or education or divide workers categories, e.g., of laborers and management (Willis, 1986). However, a growing body of literature suggests that years of training and broad worker categories fail to address the importance of skill specialisation, diversity, and recombination in knowledge generation (Hong & Page, 2004; Lazear, 2004; Woolley et al., 2010; Ren & Argote, 2011; Aggarwal & Woolley, 2013). In addition, the rise of the knowledge economy (Powell & Snellman, 2004) has sparked new interest in a more nuanced measure of skill composition. In this context, several papers have taken skill diversity and individual cognitive abilities into account for estimating their effect on wages (Bowles et al., 2001; Heckman et al., 2006; Borghans et al., 2008; Altonji, 2010; Autor & Handel, 2013).

A central conclusion of past contributions on skill diversity is that the relationship between wages and skills does not only depend on a worker's individual skills but also, how they are combined. The question of skill synergies arises (Allinson & Hayes, 1996). For some skills (e.g., programming in JavaScript and visualisation techniques) it can be argued that skill synergies emerge. The bundle of skills is more valuable than the sum of its parts. It could be argued that skill synergies are constrained to an occupational domain, e.g., programming in python and translating Russian should have little skill synergies. Certainly, the value of additional skills depends on the skill portfolio that the worker already possesses (Altonji, 2010). However, this work precisely investigates how limited synergy effects of skill bundling are and if cross-skilling, the acquisition of a new skill outside of the existing skill portfolio, might indeed be profitable.

## Method and Data

### *Online Labour Market Data*

The data for this analysis stems from the freelancing platform UpWork[4], which falls under the category of online labour markets (OLM). These platforms are websites that mediate between buyers and sellers of remotely deliverable cognitive work (Horton, 2010). The sellers of work on OLMs are either people in regular employment earning additional income by "moonlighting" via the Internet as freelancers or they are self-employed independent contractors. The buyers of work range from individuals and early-stage startups to Fortune 500 companies (Corporaal & Lehdonvirta, 2017). OLMs can be further subdivided into microtask platforms, e.g., Amazon Mechanical Turk, where payment is on a piece rate basis or freelancing platforms, such as UpWork, where payment is on an hourly or milestone basis (Lehdonvirta, 2018).

Between 2017 and 2020, the global market for online labour has grown approximately 50% (Kässi & Lehdonvirta, 2018). In light of the COVID-19 pandemic and it's significant economic repercussions across industries (Stephany et al., 2020a), OLMs continue to increase in popularity due to a general trend of work at distance (Stephany et al.. 2020b). UpWork is usually perceived as the globally most popular freelance platform (Kässi & Lehdonvirta, 2018). This study utilises OLM data, as platforms like UpWork have become early "laboratories" of the rebundling of skill sets. Their data allow us to monitor skill rebundling in a global workforce by near real-time reporting location, asking wages, previous income, gender attributes (forenames), and up-to-date skill bundles on a granular level.

[4] 4,810 freelancer profiles of a global workforce, openly accessible, have been web-scraped via the use of the general programming language Python 2.0 between October 4th and October 10th 2020.

*Method*

For the methodological approach of this paper, the work by Anderson (2017) is referential. Anderson constructs a human capital network of skills from online freelancers and shows that workers with diverse skills earn higher wages. The limitation of Anderson's work is that a skill specific evaluation in the context of cross-skilling is not addressed. This work aims at adding this cross-skilling perspective and exemplary sketches economically valuable cross-skilling pathways in times of shifting occupational profiles.

Similar to Anderson (2017), this work uses the rich toolbox of network analysis for the characterisation of skill relationships. Given a sample of 14,790 freelancers with multidimensional skill portfolios, a network is constructed in which 3,480 skills are nodes and two skills are connected by a link if a worker has both. Links are weighted according to how often the two skills co-occur. First, this skill network provides us with an endogenous categorisation of skills based on their relationship in application and the context dependency of human capital. Figure 1 illustrates this process.

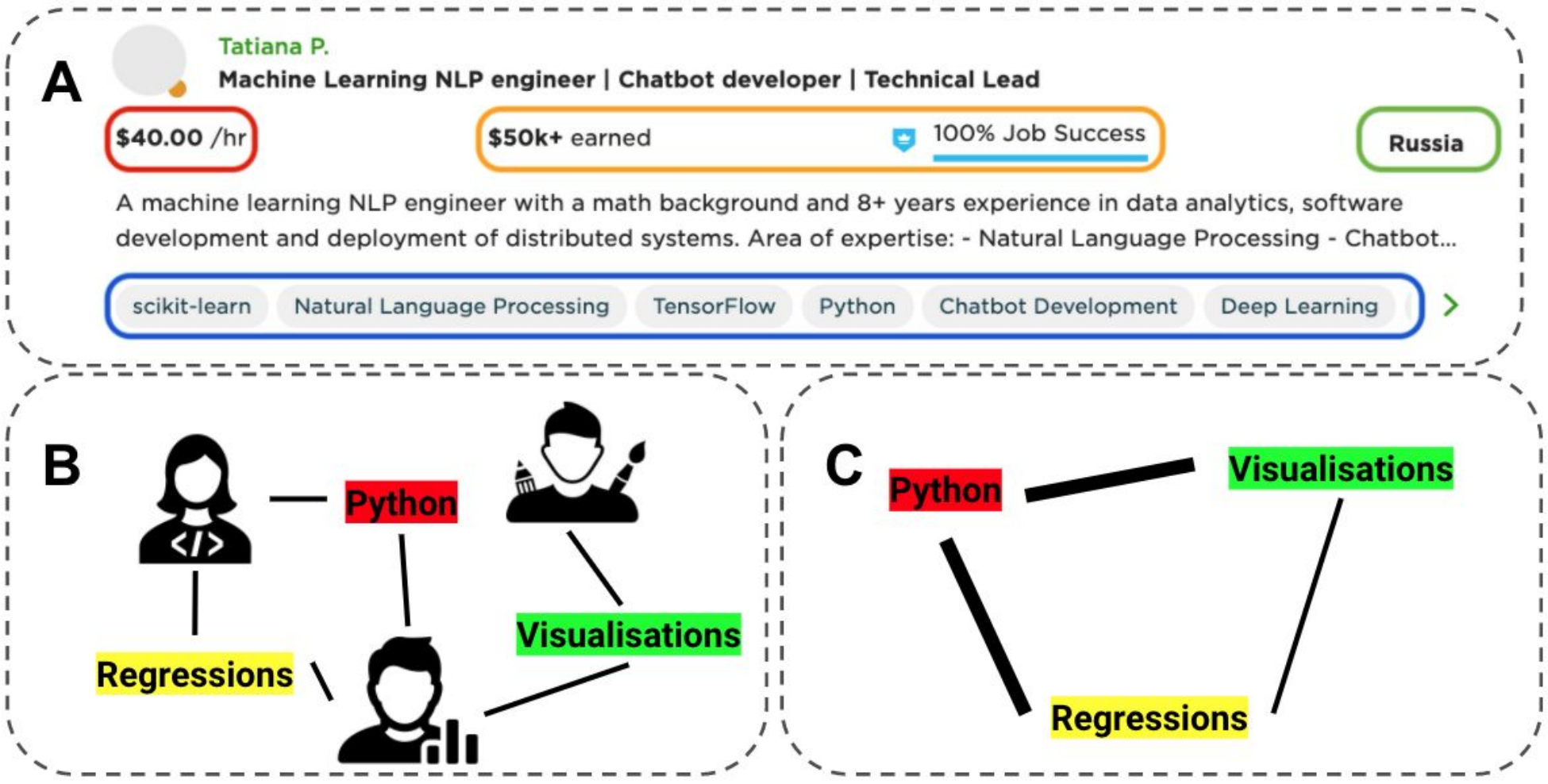

**Figure 1:** *Given a sample of 14,790 freelancers with multidimensional skill portfolios, a network is constructed in which 3,480 skills are nodes and two skills are connected by a link if a worker has both.*

In a second step, the wage proposals[5] of workers allow a statistical assessment of skills. In the given model, via calculating regression coefficients, the economic value of the 13 most popular programming languages[6] can be derived:

[5] This study relies on asking wages rather than on achieved past wages as it can be assumed that the asking wages of experienced workers resembles their actual market wage with a little margin of error. Alternatively, one could use the average of the worker's last achieved wages with the downside that these wages would rather reflect the past than the current market evaluation of a worker's skill offer. In addition, as the model controls for past earnings, the observed variation in skill coefficients should not be attributable to the past success of freelancers.

[6] Once a larger set of skills is considered at the same time, the model validity is scrutinised by potential multicollinearity.

$$wage_i = \beta_0 + \beta_1 * country_i + \beta_2 * log(earned)_i + \beta_3 * diversity_i + \beta_4 * skill_{i,j} + e_i$$
$$(1)$$
$$i \ \varepsilon \ 1, ..., n \ and j \ \varepsilon \ 1, ..., 13$$

The linear regression model (1) uses all workers ($i \ \varepsilon \ 1, ..., n$) as observations and considers their country of origin, the skill diversity of their profile[7], and amount of money earned as characteristics when estimating the worker's asking wage. In addition, each of the 13 most popular programming languages are considered as an explanatory feature in the linear regression. Lastly, the method addresses the issue of cross-skilling. Based on the existing skill portfolio of a worker, skill coefficients are again calculated. This time only a subset of workers is considered:

$$wage_k = \beta_0 + \beta_1 * country_k + \beta_2 * log(earned)_k + \beta_3 * diversity_k + \beta_4 * skill_{k,j} + e_i,$$
$$(2)$$
$$k \ \varepsilon \ a, ..., m < n \ and j \ \varepsilon \ 1, ..., 13$$

Here the accounts between *a* and *m* fall into a specific occupational domain, e.g., translation and writing, as the majority of their skills are located in this skill cluster. Hence the coefficient of the skill characteristic ($\beta_4$) only refers to the additional wage this skill contributes within a smaller subset of workers with the skill bundle $k \ \varepsilon \ a, ..., m < n$. This cross-referencing allows us to indicate the potential additional marginal value of acquiring a new skill if added to a specific skill portfolio.

## Results

### *The Clustering of Skills*

As the first part of the analysis a skill network is constructed, shown in Figure 1. The network uses the information of 14,790 workers and 3,480 unique skills. In this network, unique skills are represented as nodes. They are connected if simultaneously advertised by the same worker. The edges between two nodes grow in strength the more workers combine a pair of skills. The (unweighted) degree centrality of each node is represented by its size. Based on the relationship between skills via workers, a Louvain clustering method is applied[8] that minimises the number of edges crossing each other. Seven distinct clusters emerge, as highlighted in different colours. By highlighting the ten most prominent skills - in terms of degree centrality - of each cluster, we can see conceptual consistency within clusters and in distinction to neighbouring groups of skills. For further analysis, the eight clusters are labelled with: 3D Design (top left), Admin

[7] Skill diversity is measured by the number of occupational domains a worker's skills stem from.

[8] With the use of the network analysis software Gephi using a Force2 layout algorithm (Jacomy et al. 2014).

Support (middle right), Audio Design (bottom left), Graphic Design (middle left), Legal (bottom right), Software and Technology (top middle), Data Engineering (top right), and Translation and Writing (bottom middle).

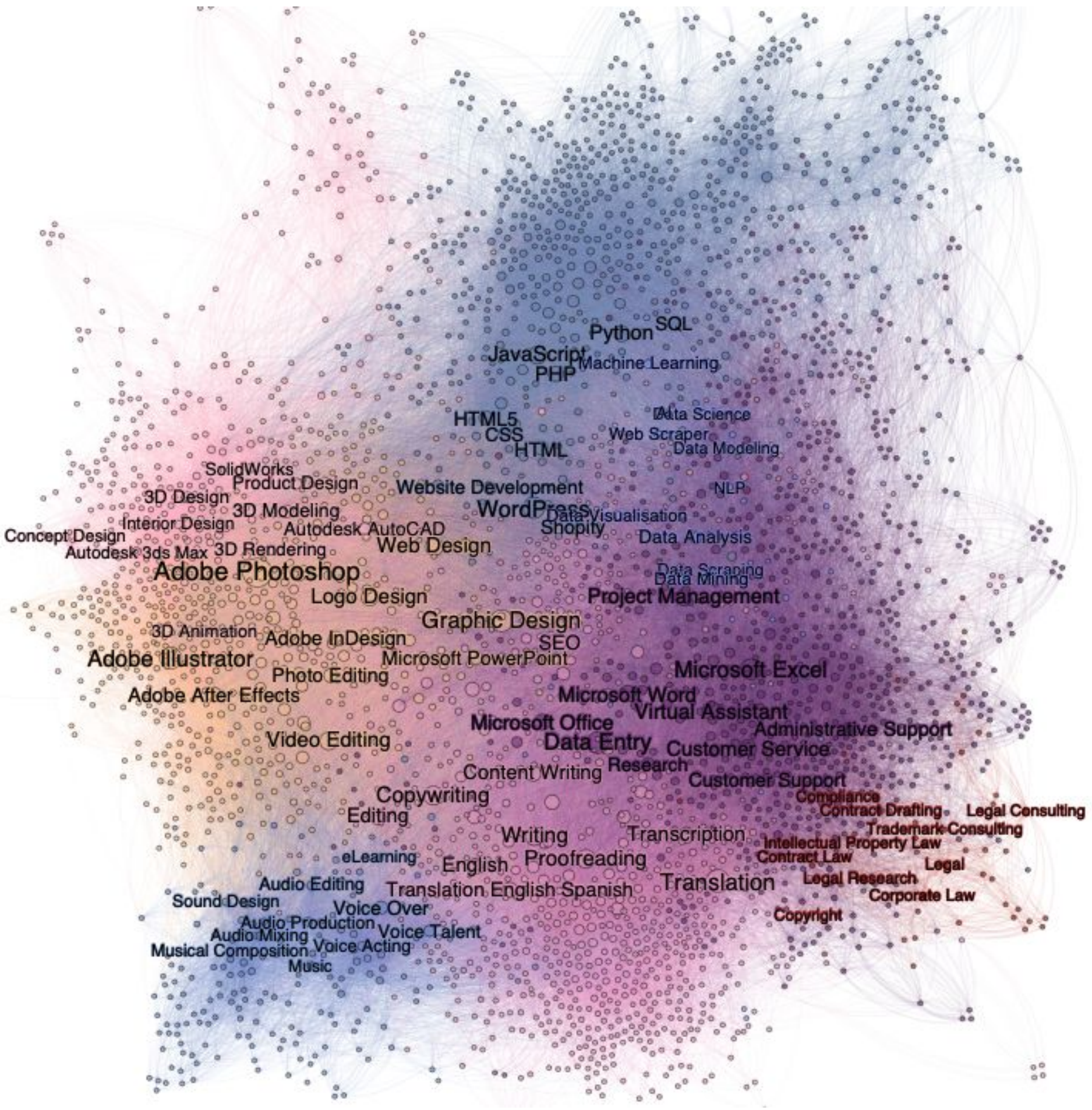

**Figure 2:** *In a skill network, skills are connected if jointly advertised by the same worker. Skills group in seven clusters with different degrees of centrality (node size). The largest skill clusters are Admin and Support, Translation and Writing, and Software and Technology.*

The skill clusters differ in size (of skills and workers employing them) but their conceptual consistency underlines the effectiveness of this endogenous clustering approach. In comparison to human classification of skills in the context of online freelance markets, similarities and differences occur. Kässi and Lehdonvirta (2018), for example, classify skills in six different domains that do not include 3D, Graphic, and Audio Design individually. The endogenous clustering approach, however, clearly indicates that these skill groups form individual clusters of their own but they are similar to each other and though group together at the left hand side of the skill network.

Skill clusters also differ significantly with regard to the asking wages of workers, as shown in Figure 3, ranging from a median of 17 USD per hour, asked by workers in Admin and Support to 33 USD/hour in Software and Tech. to 85 USD/hour in Legal. However, the wage spread within and across skill domains is sizable. In Software and Technology, asking wages range from 3.75 to 495 USD per hour.

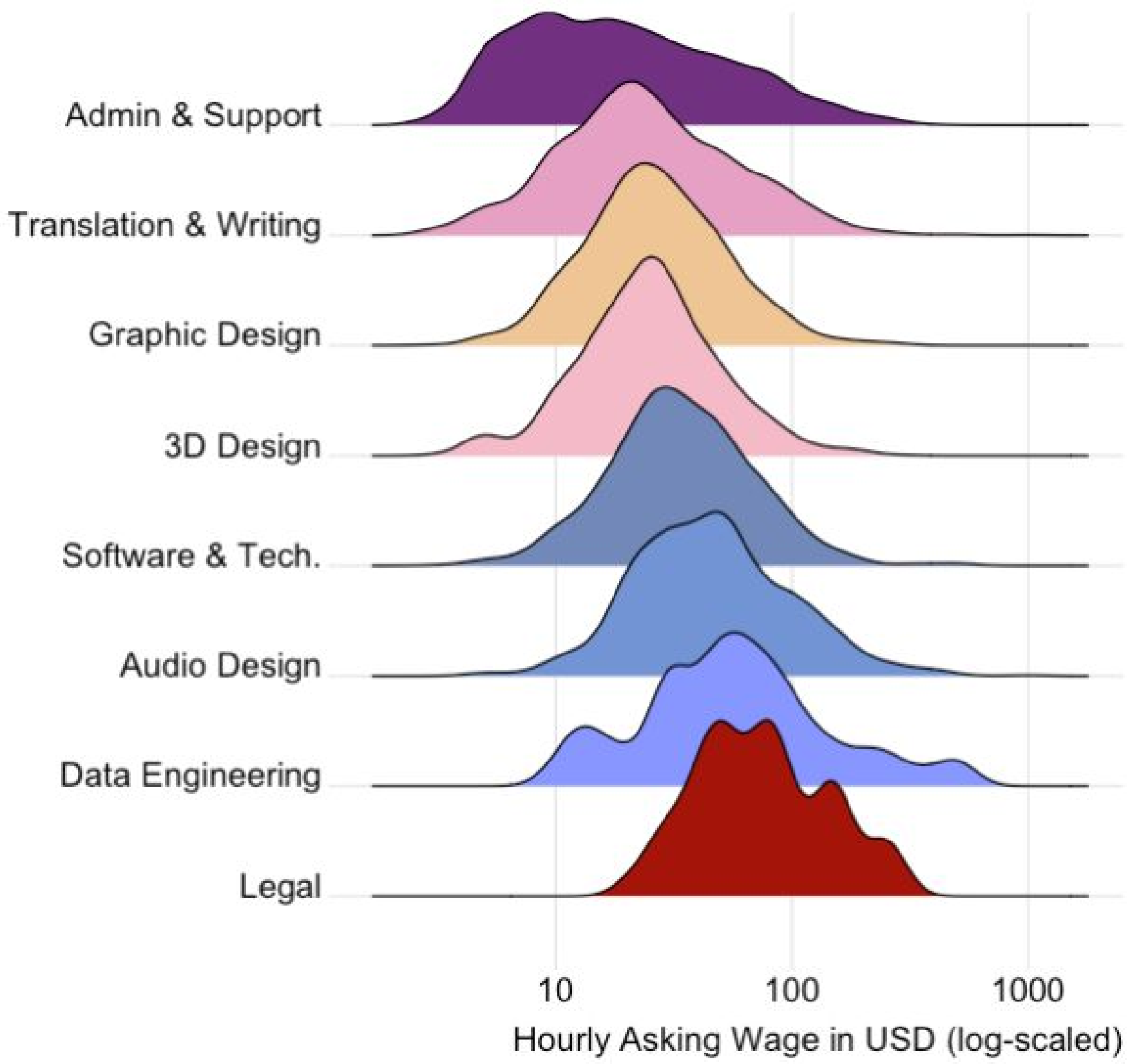

**Figure 3:** *Skill groups differ significantly with regard to asking wages. At the same time, the spread of wages within groups is sizable, too.*

## *The Value of Up-Skilling*

The rich variety of skill combinations allows us to assess the value of adding a new skill to a worker's skill portfolio. Via linear regression models, we can calculate beta coefficients for the 13 most popular programming skills for seven domains[9], as shown in Figure 4.

[9] The category of Legal with only a handful of jobs is excluded in this analysis.

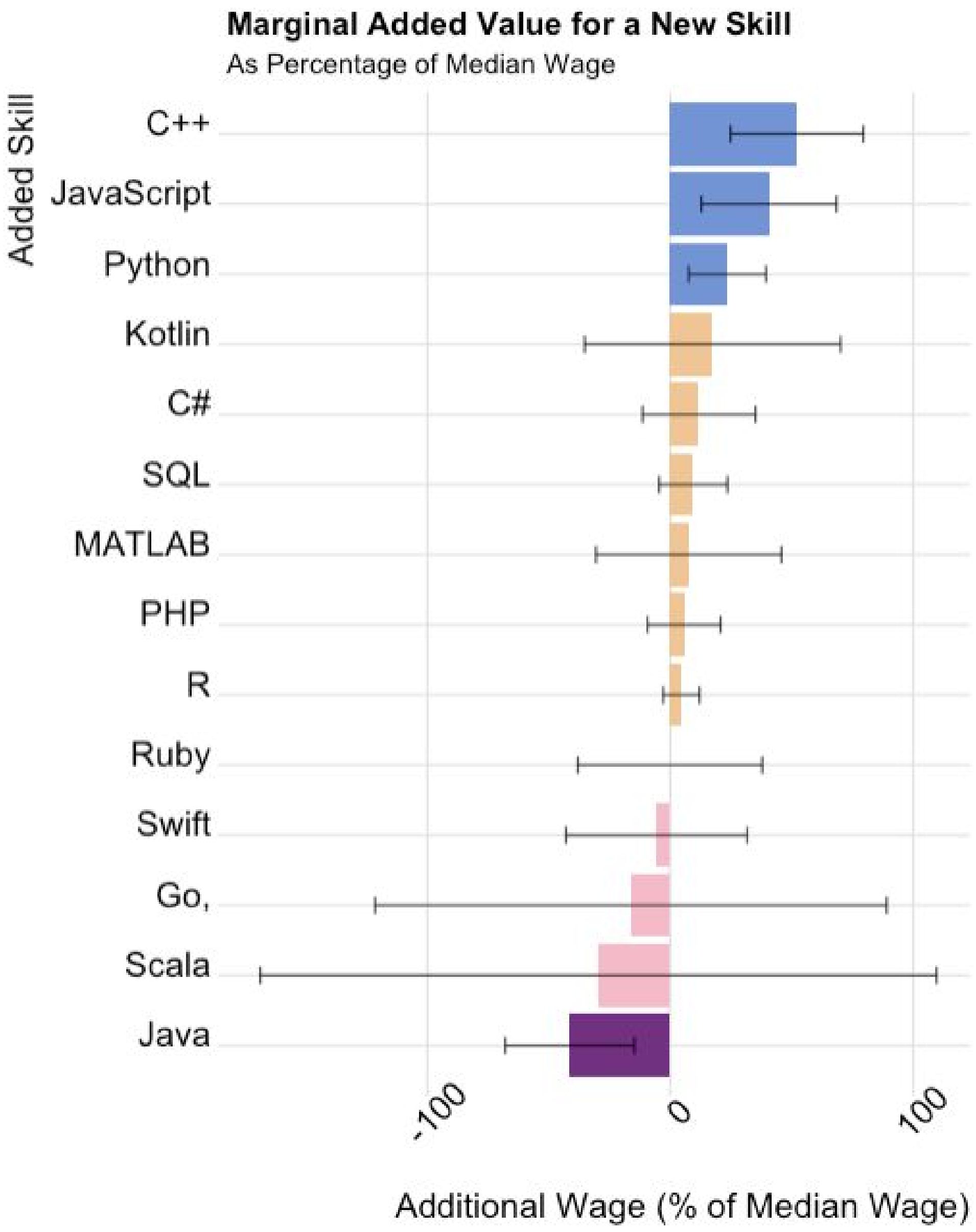

**Figure 4:** *The marginal economic benefit of learning a new skill varies significantly across the spectrum of the 13 most popular programming skills.*

The added value of learning one of the most popular 13 programming skills varies significantly. On average, for three of the skills the beta coefficient is positive, for four skills (C++, JavaScript, Python), results are statistically significant ($p>0.01$). But even within the group of significantly profitable skills, the spread is large. Being knowledgeable in C++ adds 52% to the media worker's asking wage of the whole sample, while Python skills contribute 23%[10]. Negative coefficients, as for Java, indicate that workers with these skills ask for significantly lower wages than the average online freelancer.

[10] Precise values and the coefficients for other features can be found in the Appendix.

*The Complementary Value of Skills*

With the endogenous classification of skill groups at hand, we can perform an evaluation of learning a new skill based on the already existing skill bundle of a worker. For this purpose, two analyses are performed. First, the value of skill diversity is assessed. Skill domains are attributed to each worker based on her skill bundle. While each worker has a dominant skill category (the relative majority of all of her skills are in this domain), some workers also add skills from other domains to their portfolio. In Figure 5, wage distributions of different skill diversities are shown across the major six skill groups. In general, it can be noticed that across skill domains, workers with more diverse skill bundles have higher wages. In particular, for workers with very diverse bundles, i.e., adding skills from three domains other than the major skill category, wages on the 4th quartile (lines) of the distribution are shifted upwards.

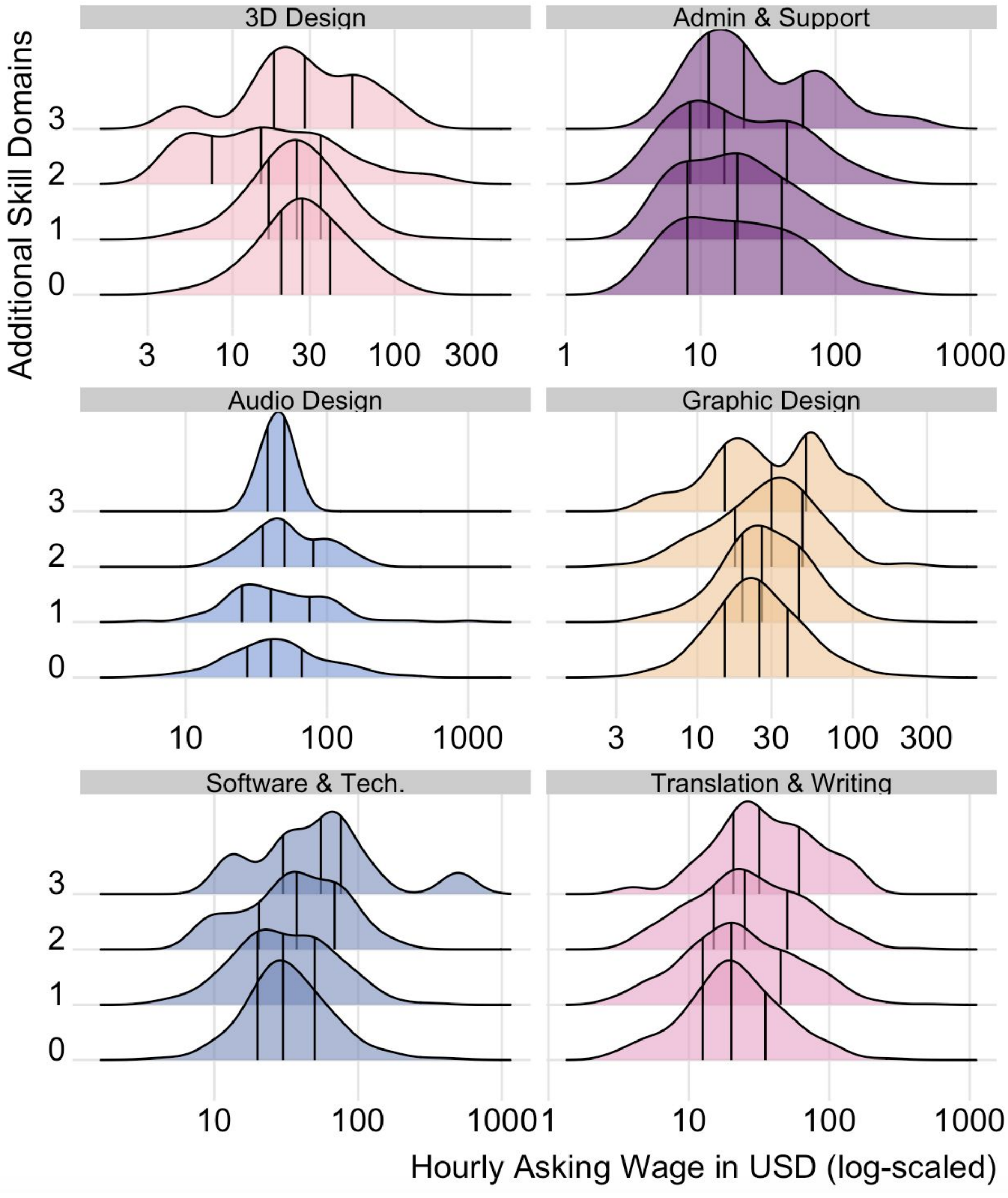

**Figure 5:** *Across skill domains, asking wages increase when skills from other domains are added to the workers' portfolio. Workers in the top earning quartiles (lines) ask for significantly higher wages when demanding skills from three domains other than their defining skill bundle (Observation sizes in Legal and Data Engineering are too small).*

In a second step, the linear regression explaining workers' asking wages is performed for the complete set of workers and the seven major skill domain subsets. Figure 6 summarises the coefficients of the programming skills in the eight scenarios.

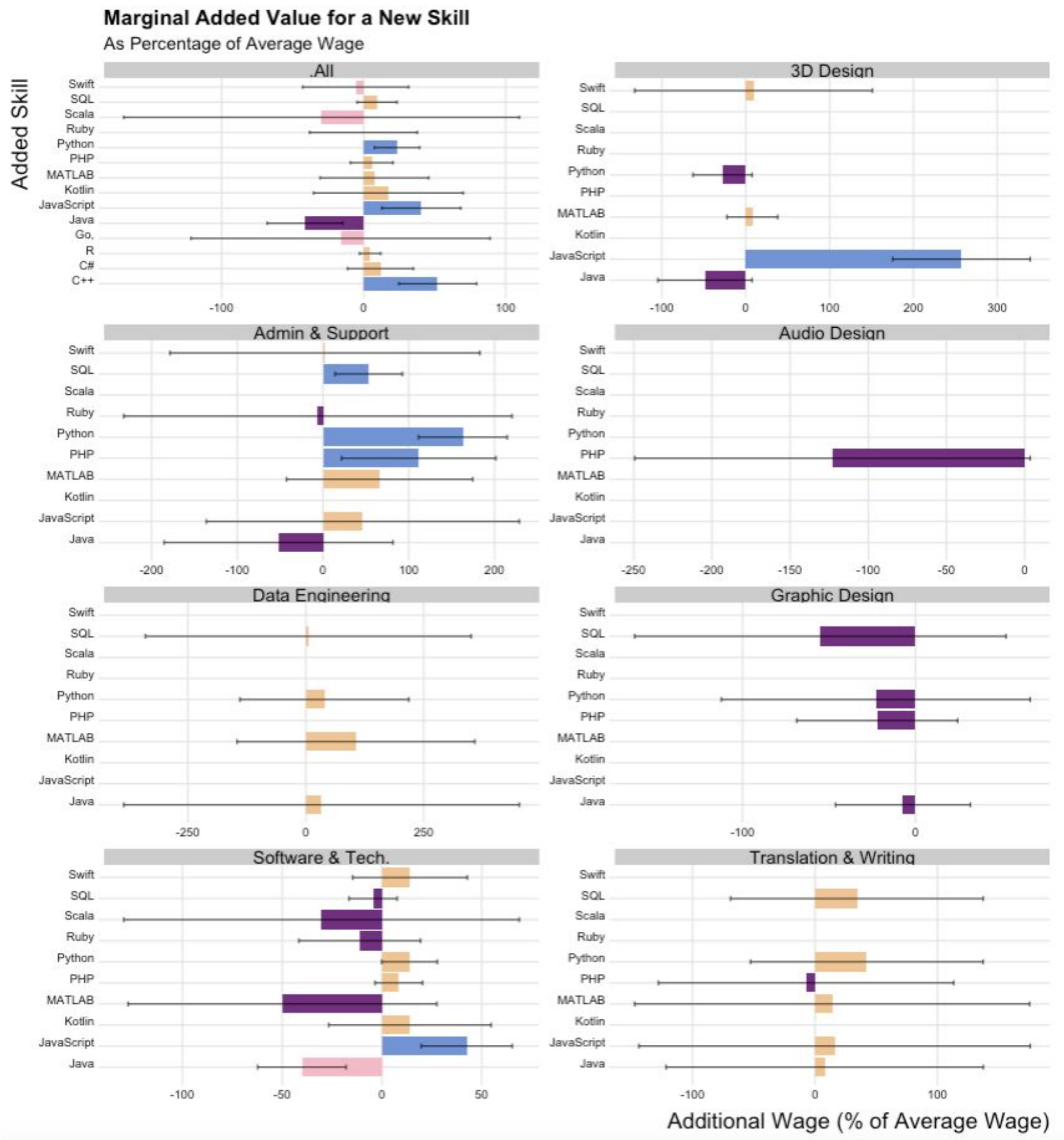

**Figure 6:** *Learning a new programming skill, like Python, pays off in general, but even more when added to skill bundles like Admin and Support.*

In contrast to the added economic value for the complete set of workers, we see that some skills, like programming in Python, increase the worker asking wage over proportionally in the skill context of Admin & Support. Similarly, knowing how to work with JavaScript contributes significantly more when added to 3D Design. Figure 7 illustrates the complementarity of skill benefits with the example of Python and Java.

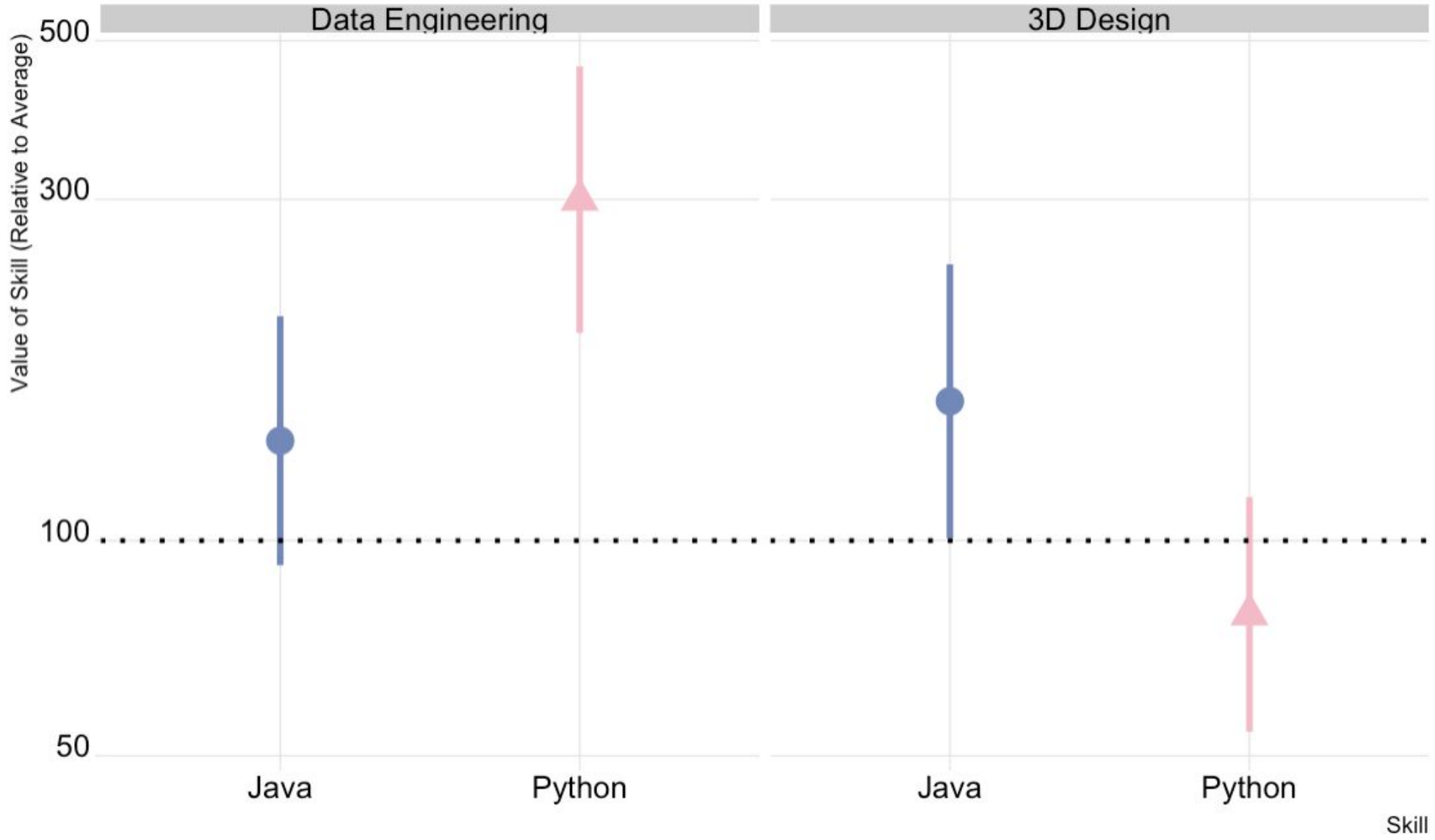

**Figure 7:** *Designers that know Python can't add wage but workers that are skilled in Java can increase their wage by more than 100% on average.*

In the field of Data Engineering it is of little surprise that knowing how to program with Python, allegedly THE data science super skill (Grus 2019), adds more to a worker wage than knowing how to work with Java. However, once we add these skills to a bundle in the domain of 3D Design, the picture turns upside down, as shown in Figure 7. Designers that know Python can't add wage but workers that are skilled in Java can increase their wage by more than 100% on average. These skill trajectories are an illustration of what online labour market data allow us to say about the complementarities of learning a new skill. The data enable us to evaluate the economic benefit of individual skills based on the existing skill bundle of a person to ultimately sketch individual re-skilling pathways.

## *Revealing the AI Skill Domain*

A further obstacle in developing effective and timely reskilling pathways is the lack of skill contextualisation of newly emerging technologies. The rapid expansion of such emerging digital technologies, like AI, is creating a huge demand for labour skilled in the development and application of these domains. However, the fast change of skill profiles in new technology environments makes it difficult for companies to find adequately trained experts. At the same time, it is unclear which types of skills constitute newly emerging areas of digital technologies (De Mauro et al. 2018).

Companies are not able to satisfy their rapidly growing demand for talents in information and communication technologies (ICT). Projections show that positions for digital technology talents  are the fastest growing job segment in the United States with estimated 2,720,000 openings by the end of 2020 (Miller and Hughes 2017), leading to a significant excess demand for ICT professionals. The results of qualification mismatches are lower labour and economic productivity (McGowan & Andrews 2015). This can be particularly harmful for economies in regions with low levels of growth, where expert labour in ICT is already scarce.

Often, the description of skills and responsibilities of experts in domains such as Big Data or AI is fuzzy and firms tend to apply subjective interpretations. As De Mauro et al. (2018) illustrate in detail, the emergence of the expert role of the Data Scientist is a typical example for a simplistic job description that downsizes the complexity and variety of skills required to retrieve information and transform it into economically valuable insights. Research indicates that there is a clear gap regarding the formal taxonomy of skills and educational needs in new technology domains like AI (Miller & Hughes 2017; Song & Zhu 2016).

The presented approach of creating skill networks can help to reveal precise skill sets that are related to AI. Figure 6 shows a subset of nodes from the skill network in Figure 1. Here, only skills from profiles are considered that have appeared under the search term "Artificial Intelligence". Similar to Figure 1, skills are connected if jointly advertised by the same worker.

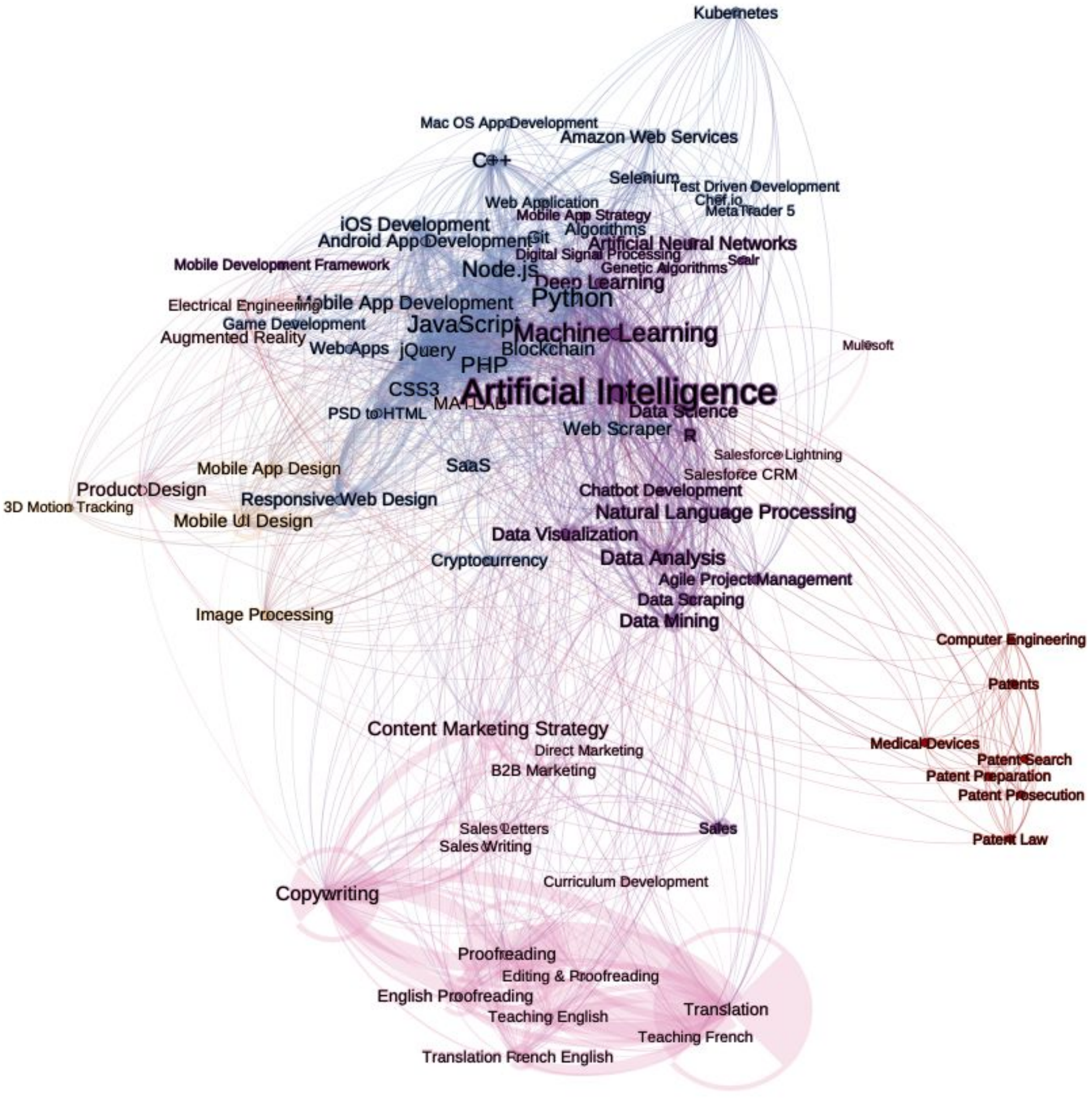

**Figure 7:** *This network is a subset of Figure 1. Only skills that have been advertised in the context of the search term "Artificial Intelligence" are selected. The skills related to AI are strongly concentrated in the cluster of software and technology.*

Skills are strongly clustered around the domains of software and technology and admin and support. In absolute terms, AI appears most frequently in the skill context of software and technology; 28 skills related to AI belong to this domain. However, taken the size of the different skill clusters into account, legal is the skill domain most strongly populated by AI skills; seven out of 46 legal skills (15.2%) are connected to AI.

## Conclusion and Discussion

### *Sketching Cross-Skilling Pathways*

As technological change accelerates, task automation shifts occupational skills requirements, challenging the global workforce to constantly re-skill. To avoid skill gaps and systematic labour market mismatches, approaches to reskilling need to step-up, as traditional education policies are too slow for the fast-changing pace of technological and social change. In addition, situations like the COVID-19 lockdown further accelerate digitalisation trends while limiting economic resources of companies to up-skill their employees and constraining workers to learn remotely from home.

In light of this grand challenge, this work explores the foundations of new modes of re-skilling via sketching cross-skilling pathways based on online labour market data. Online labour markets have become early laboratories for the de- and rebundling of skills from previously unrelated domains. The statistical analysis of diverse skill portfolios and wages of online workers allows an evaluation of the economic benefit of learning a new skill. Furthermore, the endogenous categorisation of skills via skill networks gives us insights into the value of learning a new skill depending on the already existing skill portfolio of each worker. We see that some skills are, in terms of additional wage, more valuable than others.

In addition to the economic evaluation of individual skills, this work assesses the added economic value of learning a new skill complementing an existing skill portfolio. The conditioning on skill domains is a relevant perspective, as individual examples show. Mimicking the individual cost-benefit analysis of learning a new skill is mandatory for future educational formats, as skill portfolios become more fragmented and re-skilling opportunities more granular.

### *Limitations and Future Work*

This work is a first exploration of a quantitative and market data based assessment of complementarities in learning a new skill. It comes with limitations and opens space for future investigations. The strong point of this work is that it measures the supply side of the labour market and allows skill evaluations via wages that are complementary to the existing skill bundle of the worker. However, these claims are currently based on the analysis of observational and cross-sectional data. Models of statistical inference are constructed to isolate the contribution of an individual skill to a worker's wage, in addition to other confounding characteristics, such as country of origin or past earnings. While significant economic benefits of single skills are made visible, the current model is weak on establishing causal linkages between acquiring a new skill and a change in worker wages. Some, potentially relevant, confounding worker characteristics, such as a worker's attitude or communication style, can not be observed. Furthermore, in a setting with observational and cross-sectional data, wage differences are aggregations of a general population of freelancers. In contrast, following the history of individual workers over time and observing their wage before and after the actual acquisition of a new skill would allow us to overcome these limitations and have a closer look at causal linkages between learning a new skill, workers' success, and earnings.

# Appendix

**Table 1:** *Network metrics of the 35 skills with the highest degree centrality in each of the seven main skill cluster*

| Skill | Degree | Group |
|---|---|---|
| 3D Modeling | 327 | 3D Design |
| Autodesk AutoCAD | 301 | 3D Design |
| Product Design | 273 | 3D Design |
| 3D Design | 253 | 3D Design |
| 3D Rendering | 235 | 3D Design |
| Data Entry | 678 | Admin Support |
| Microsoft Excel | 639 | Admin Support |
| Virtual Assistant | 521 | Admin Support |
| Project Management | 472 | Admin Support |
| Customer Service | 462 | Admin Support |
| Voice Over | 341 | Audio Design |
| Voice Talent | 263 | Audio Design |
| Audio Editing | 214 | Audio Design |
| Audio Production | 178 | Audio Design |
| Voice Acting | 176 | Audio Design |
| Adobe Photoshop | 957 | Graphic Design |
| Graphic Design | 681 | Graphic Design |
| Adobe Illustrator | 672 | Graphic Design |
| Video Editing | 516 | Graphic Design |
| Web Design | 515 | Graphic Design |
| Corporate Law | 55 | Legal |
| Legal Research | 54 | Legal |
| Contract Law | 53 | Legal |
| Contract Drafting | 49 | Legal |
| Legal | 48 | Legal |
| WordPress | 660 | Software and Tech. |
| JavaScript | 516 | Software and Tech. |
| Python | 442 | Software and Tech. |
| PHP | 423 | Software and Tech. |
| HTML | 368 | Software and Tech. |
| Translation | 715 | Translation and Writing |
| Copywriting | 564 | Translation and Writing |
| Writing | 558 | Translation and Writing |
| Proofreading | 542 | Translation and Writing |
| Transcription | 447 | Translation and Writing |

**Table 2:** *Summary of the linear regression, earnings, skill diversity, and skill on workers' asking wages (only significant values and no country dummies are shown).*

| | *Dependent variable:* |
|---|---|
| | wage |
| Python | 5.882*** |
| | (2.031) |
| Java | −10.347*** |
| | (3.398) |
| JavaScript | 10.157*** |
| | (3.546) |
| C++ | 13.028*** |
| | (3.491) |
| C# | 2.961 |
| | (2.962) |
| Kotlin | 4.364 |
| | (6.720) |
| Swift | −1.398 |
| | (4.759) |
| SQL | 2.371 |
| | (1.793) |
| R | 1.147 |
| | (0.954) |
| PHP | 1.415 |
| | (1.916) |
| MATLAB | 1.902 |
| | (4.879) |
| Scala | −7.400 |
| | (17.770) |
| Go | −4.063 |
| | (13.439) |
| Ruby | −0.032 |
| | (4.842) |
| Admin & Support | 3.765*** |
| | (1.438) |
| Audio Design | 17.190*** |
| | (1.707) |
| Data Engineering | 44.881*** |
| | (5.208) |
| Graphic Design | 1.825 |
| | (1.278) |
| Legal | 38.620*** |
| | (6.064) |
| Software & Tech. | 8.701*** |
| | (1.840) |
| Translation & Writing | 0.698 |
| | (1.307) |
| Earnings (log) | 3.036*** |
| | (0.092) |
| One Domain | 28.404*** |
| | (8.029) |
| Two Domains | 6.290 |
| | (4.814) |
| Three Domains | 8.111* |
| | (4.816) |
| Four Domains | 8.814* |
| | (4.842) |
| Five+ Domains | 15.451*** |
| | (5.005) |
| Constant | −11.760* |
| | (6.823) |
| Observations | 13,304 |
| $R^2$ | 0.308 |
| Adjusted $R^2$ | 0.300 |
| Residual Std. Error | 35.058 (df = 13143) |
| F Statistic | 36.624*** (df = 160; 13143) |